\documentclass[sigconf, nonacm]{acmart}
\usepackage{graphicx}
\usepackage{balance}  
\usepackage{epsfig}
\usepackage{graphicx}
\usepackage{epstopdf}
\usepackage{latexsym}
\PassOptionsToPackage{noend}{algpseudocode}
\usepackage{algpseudocode}
\usepackage{pifont}
\usepackage{epsfig}
\usepackage{amssymb}
\usepackage{amsmath}
\usepackage{amsfonts}
\usepackage{subfigure}
\usepackage{stmaryrd}
\usepackage{url}
\usepackage{multirow}
\usepackage{array}
\usepackage[normalem]{ulem}
\usepackage{color}
\usepackage{cleveref}
\usepackage{xspace}
\usepackage{mathtools}
\usepackage{soul}
\usepackage{listings}
\usepackage{enumitem}
\usepackage{xcolor}
\usepackage{tikz}
\usepackage{braket}
\newsavebox{\blackball}
\newsavebox{\greenball}

\usepackage[utf8]{inputenc}
\usepackage{enumitem}

\usepackage{bbding}
\usepackage{amssymb}
\usepackage{epstopdf}
\usepackage{epsfig,endnotes}
\usepackage{url}
\usepackage{array}
\usepackage{booktabs}
\usepackage{threeparttable}
\usepackage{tabulary}

\usepackage{CJKutf8}
\usepackage[most,breakable]{tcolorbox}
\usepackage{etoolbox}
\usepackage{fancyhdr}
\usepackage{lipsum}
\usepackage[fencedCode]{markdown}
\usepackage[table,xcdraw]{xcolor}
\usepackage{tabularx}
\usepackage{booktabs}
\usepackage{tabularx}
\usepackage{array}
\usepackage{makecell}

\newcolumntype{L}[1]{>{\raggedright\arraybackslash}p{#1}}
\newcolumntype{Y}{>{\raggedright\arraybackslash}X}

\usepackage[lined,boxed,vlined,ruled,linesnumbered]{algorithm2e}

\usepackage{bm}

\theoremstyle{definition}

\newcolumntype{M}[1]{>{\centering\arraybackslash}m{#1}}
\errorcontextlines\maxdimen

\newcommand{\sys}{\textsc{LBO}\xspace}

\newcommand{\hi}[1]{\vspace{.25em} \noindent {\bf #1}\xspace}

 \newcommand{\bfit}[1]{\textbf{\textit{#1}}}

\newcommand{\zxh}[1]{\textcolor{red}{#1}}
\newcommand{\zw}[1]{\textcolor{purple}{#1}}

\definecolor{codegreen}{rgb}{0,0.6,0}
\definecolor{codegray}{rgb}{0.5,0.5,0.5}
\definecolor{codepurple}{rgb}{0.58,0,0.82}
\definecolor{backcolour}{rgb}{0.95,0.95,0.92}

\lstdefinestyle{mystyle}{
	backgroundcolor=\color{backcolour},   
	commentstyle=\color{codegreen},
	keywordstyle=\color{magenta},
	numberstyle=\tiny\color{codegray},
	stringstyle=\color{codepurple},
	basicstyle=\ttfamily\footnotesize,
	breakatwhitespace=false,         
	breaklines=true,                 
	captionpos=b,                    
	keepspaces=true,                 
	numbers=left,                    
	numbersep=5pt,                  
	showspaces=false,                
	showstringspaces=false,
	showtabs=false,                  
	tabsize=2
}
\lstdefinestyle{jsonStyle}{
	basicstyle=\small\ttfamily,
	columns=fullflexible,
	showstringspaces=false,
	commentstyle=\color{codegreen}\upshape,
	stringstyle=\color{codegreen},
	morestring=[b]",
	moredelim=[s][\color{codepurple}]{\{}{\}},
	moredelim=[s][\color{codepurple}]{[}{]},
	moredelim=[l][\color{codepurple}]{:},
	moredelim=[l][\color{codepurple}]{,}
}

\newcommand{\oursys}{\texttt{Qute}\xspace} 

\begin{document}

\title{Hybrid Quantum Database [Vision]}
\title{\oursys: Towards Quantum-Native Database}

\author{Muzhi Chen}
\affiliation{
    Shanghai Jiao Tong University
    \city{}
    \country{}
}
\email{inefable@sjtu.edu.cn}

\author{Xuanhe Zhou}\authornote{Xuanhe Zhou is the corresponding author.}
\affiliation{
    Shanghai Jiao Tong University
    \city{}
    \country{}
}
\email{zhouxuanhe@sjtu.edu.cn}

\author{Wei Zhou}
\affiliation{
    Shanghai Jiao Tong University
    \city{}
    \country{}
}
\email{wzdb@sjtu.edu.cn}

\author{Bangrui Xu}
\affiliation{
    Shanghai Jiao Tong University
    \city{}
    \country{}
}
\email{dreameter@sjtu.edu.cn}

\author{Surui Tang}
\affiliation{
    Shanghai Jiao Tong University
    \city{}
    \country{}
}
\email{tangsurui@sjtu.edu.cn}

\author{Guoliang Li}
\affiliation{
    Tsinghua University
    \city{}
    \country{liguoliang@}
}
\email{tsinghua.edu.cn}

\author{Bingsheng He}
\affiliation{
    National University of Singapore
    \city{}
    \country{}
}
\email{dcsheb@nus.edu.sg}

\author{Yeye He}
\affiliation{
    Microsoft Corporation
    \city{}
    \country{}
}
\email{yeyehe@microsoft.com}

\author{Yitong Song}
\affiliation{
  \institution{Hong Kong Baptist University}
  \country{}
  }
\email{ytsong@hkbu.edu.hk}

\author{Fan Wu}
\affiliation{
    Shanghai Jiao Tong University
    \city{}
    \country{}
}
\email{fwu@cs.sjtu.edu.cn}

\pagestyle{plain}
\pagenumbering{arabic}

\begin{abstract}
\begin{sloppypar}

This paper envisions a quantum database (\oursys) that treats quantum computation as a first-class execution option. Unlike prior simulation-based methods that either run quantum algorithms on classical machines or adapt existing databases for quantum simulation, \oursys instead (i) compiles an extended form of SQL into gate-efficient quantum circuits, (ii) employs a hybrid optimizer to dynamically select between quantum and classical execution plans, (iii) introduces selective quantum indexing, and (iv) designs fidelity-preserving storage to mitigate current qubit constraints. We also present a three-stage evolution roadmap toward quantum-native database. Finally, by deploying \oursys on a real quantum processor (origin\_wukong), we show that it outperforms a classical baseline at scale, and we release an open-source prototype at {\it \textcolor{gray}{\url{https://github.com/weAIDB/Qute}}}.

\end{sloppypar}
\end{abstract}

\maketitle

\vspace{-.75em}
\section{Introduction}
\label{sec:introduction}






Quantum databases aim to leverage quantum computers to accelerate workloads that are increasingly infeasible for classical machines. For example, analytical query processing in columnar OLAP systems often depends on linear scans over unindexed tables~\cite{DBLP:journals/pvldb/ZengHSPMZ23}. By leveraging entangled qubit superpositions and quantum algorithms such as Grover's search~\cite{quandbsurvey}, quantum databases can substantially accelerate these queries. Recent research follows two main directions. First, \textit{{quantum computation for database}} applies quantum algorithms to classical databases for tasks like query optimization~\cite{queryoptimization1,liu2025q2o}, index tuning~\cite{indexselection1}, approximate query processing~\cite{quanmeetdb}. However, they remain simulation-based and lack full-stack quantum integration. Second, \textit{{database for quantum computation}} leverages classical database techniques to manage quantum workloads, such as encoding quantum circuits in tabular form, but does not consider optimization on the database side~\cite{qudatadiscuss,quanviable}.


\begin{sloppypar}
To address these limitations, we pursue an end‑to‑end quantum‑native database, which treats quantum computation not as an isolated accelerator but as a first‑class execution substrate integrated throughout query parsing, planning, and execution. However, there are four main challenges. First, SQL operators must be reformulated into representations that map naturally to quantum primitives (e.g., superposition, interference). Second, the optimizer must plan across noise‑sensitive quantum operators whose latency, success probability, and approximation error differ fundamentally from classical optimizers. Third, 
the limited qubit budget makes it necessary to selectively probe only a small fraction of the dataset during quantum execution.
Finally, quantum-active attributes must be stored in fidelity-preserving formats to maintain high usability.
\end{sloppypar}

\begin{figure}[!t]
  \centering
  \includegraphics[width=\linewidth]{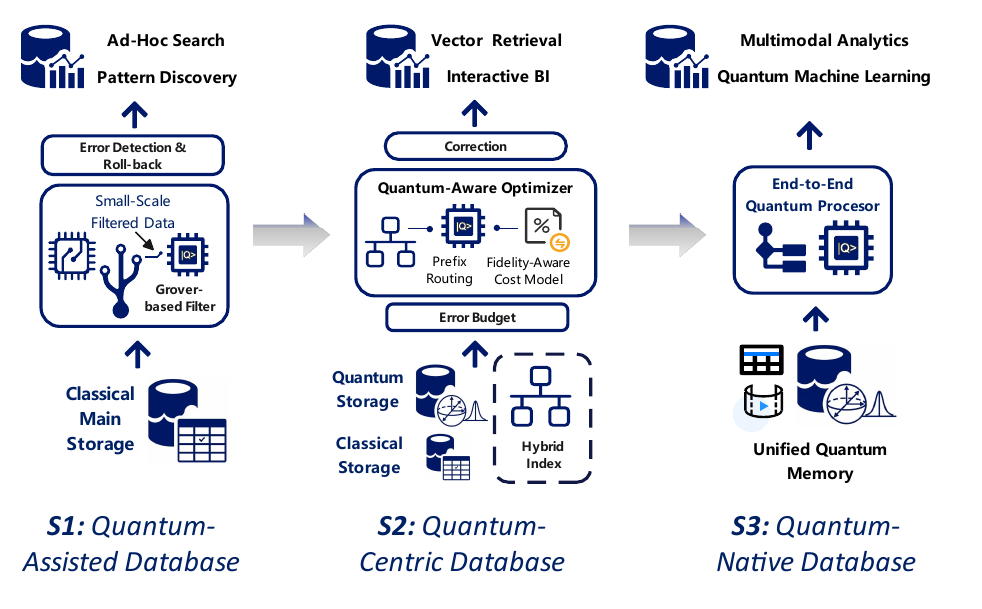}
  \vspace{-.75cm}
  \caption{Potential Evolution of Quantum Databases.}
  \label{fig:intro}
  \vspace{-.75cm}
\end{figure}


In this paper, we introduce \oursys, an integrated quantum database architecture that addresses the above challenges. First, \oursys compiles extended SQL queries into quantum-gate efficient circuits via a three-tier intermediate representation. Second, it employs a hybrid optimizer that combines quantum and classical operators. Third, it offloads costly operators (e.g., similarity joins) onto quantum circuits that encode computation into probability amplitudes. Fourth, to address scale limitations like qubit numbers, \oursys introduces selective quantum indexing based on hybrid tree structures enhanced. Fifth, it adopts a fidelity-aware data format and storage engine supporting compressed tensor networks.  


As shown in Figure~\ref{fig:intro}, we envision the development of quantum-native databases across three stages. In the first stage, quantum computing serves as a co-processor for classical systems, where data remains in classical memory and only small subsets are loaded into quantum registers for basic tasks like data filtering. This stage lacks index acceleration and relies on rollback mechanisms to maintain correctness under noise. In the second stage, more data can be stored in quantum memory, enabling hybrid classical–quantum indexing and quantum operators such as joins and aggregations. In the third stage, with scalable fault-tolerant hardware, all data (including multimodal formats like video) can reside in quantum memory, enabling fully quantum-native storage, indexing, processing. 

\section{Preliminary}
\label{sec:quantum_basics}



We present quantum concepts supporting \oursys, including: (1) qubits, (2) quantum gates, and (3) quantum circuits. 

\hi{Qubit.} Classical computers encode information as deterministic bits in $\{0, 1\}$, and solve computation tasks by explicitly enumerating and evaluating candidate solutions one by one. In contrast, quantum computers encode information using \emph{qubits}, whose states are not restricted to a single deterministic value but can exist as \emph{superpositions} of the classical basis states $\ket{0}$ and $\ket{1}$. Upon \emph{measurement}, a qubit collapses to a definite value of 0 or 1~\cite{singlequbit1}. Formally, a qubit state is $\ket{\psi} = \alpha \ket{0} + \beta \ket{1}$, where amplitudes $\alpha$ and $\beta$ encode both magnitude and phase, and measurement yields 0 or 1 with probabilities $|\alpha|^2$ and $|\beta|^2$ satisfying $|\alpha|^2+|\beta|^2=1$. With qubits, numerous candidate solutions can be represented implicitly~\cite{largecandidate}, and appropriate quantum gates and circuits can amplify the probability of desired solutions, thereby reducing search complexity~\cite{datasearch2}.

\hi{Quantum Gate.} A quantum gate is an operation that changes qubit states in a controlled way, enabling amplifying correct possibilities while suppressing incorrect ones. Quantum gates are commonly classified into two types: \textit{single-qubit gates} and \textit{multi-qubit gates}.

\noindent $\bullet$ \bfit{(1) Single-qubit gate} manipulates the amplitude (e.g., $X$-axis and $Y$-axis rotations and flips) and phase (e.g., $Z$-axis rotations) of an individual qubit to achieve operations essential for data representation and quantum computation. \uline{For state preparation}, $(i)$ the \textit{Hadamard (H) gate} offers a uniform initial representation for quantum computation by mapping a basis state into an equal superposition of $\ket{0}$ and $\ket{1}$; $(ii)$ \textit{X-axis and Y-axis rotations} (i.e., $R_X(\theta)$ and $R_Y(\theta)$) encode row values by adjusting the amplitude contributions of a qubit between $\ket{0}$ and $\ket{1}$; $(iii)$ the \textit{Pauli-X flip} operates a qubit between $\ket{0}$ and $\ket{1}$, which is used to mark specific basis states. \uline{For phase adjustment}, \textit{Z-axis rotation gate} (i.e., $R_Z(\theta)$) modifies the relative phase between $\ket{0}$ and $\ket{1}$, which determines whether different possibilities reinforce or cancel each other (thereby amplifying the probability of target rows). \uline{For value encoding}, \textit{parameterized rotations} $R_\alpha(\theta)$, where $\alpha \in \{X, Y, Z\}$, adjust both magnitudes and phases, allowing {numerical} information to be embedded~\cite{singlequbit1, singlequbit2}. 

\noindent $\bullet$ \bfit{(2) Multi-qubit gate} performs conditional operations and create \emph{entanglement} among qubits, thereby establishing their correlations. Such correlations are essential for quantum parallelism and computational speedups that cannot be achieved on classical computers. Multi-qubit gates can be broadly classified into three categories: $(i)$ \textit{Controlled gates} act on target qubits only when the control qubits are in specific states; $(ii)$ \textit{Entangling gates} bind multiple qubits together to create non-classical correlations; $(iii)$ \textit{Permutation gates} exchange the states of qubits to rearrange data. For example, 
\textit{Controlled-SWAP (CSWAP) gate}, which combines controlled and permutation operations, swaps the states of two target qubits only when the control qubit is in a specific state~\cite{multiqubitgate1}.


\hi{Quantum Circuit.} 
A quantum circuit organizes a sequence of quantum gates to implement quantum algorithms and produce the desired results. We use a representative quantum circuit (i.e., \emph{Grover’s Algorithm}~\cite{Grover}) to illustrate its potential for accelerating data filtering. Given a boolean predicate $f : \mathcal{X} \rightarrow \{0, 1\}$ (in the SQL \texttt{WHERE} clause) over a dataset of size $N = |\mathcal{X}|$, data filtering identifies all records $x \in \mathcal{X}$ such that $f(x)=1$. The circuit begins by uniformly encoding the full dataset via the H gate, and then organizes a sequence of gates with the following two operators.

\noindent $\bullet$ \bfit{(1) Oracle ($\boldsymbol{O_f}$):} 
The oracle acts as a predicate evaluator, implemented as a quantum operator that encodes the query condition (e.g., $f(x) = 1$). Applied to a superposition ${\ket{x}}$, it marks records satisfying the predicate by a phase inversion, i.e., $O_f \ket{x} = (-1)^{f(x)}\ket{x}$.

\noindent $\bullet$ \bfit{(2) Diffusion Operator ($\boldsymbol{D}$):}
This operator amplifies success probability by reflecting probability amplitudes about their mean, increasing the amplitudes of marked states and suppressing others.
It is implemented by applying H gates to all qubits, performing a multi-qubit phase inversion on $\ket{0 \cdots 0}$ state, and then reversing the Hadamard transformation, i.e., ${D = 2\ket{s}\bra{s} - I}$.

By iteratively applying the operator $G = D O_f$ for $k \approx \frac{\pi}{4}\sqrt{N/M}$ steps (where $M$ is the matched record count), measurement yields the matched record with high probability. 
Unlike a classical linear scan requiring $O({N})$ predicate evaluations, the circuit finds matching records using $O(\sqrt{N})$ evaluations {(i.e., the worst case $M=1$)}, highlighting its advantage for filtering over unindexed data.

\section{System Overview}
\label{sec:overview}

\begin{figure}[!t]
  \vspace{-.25cm}
  \centering
  \includegraphics[width=.9\linewidth]{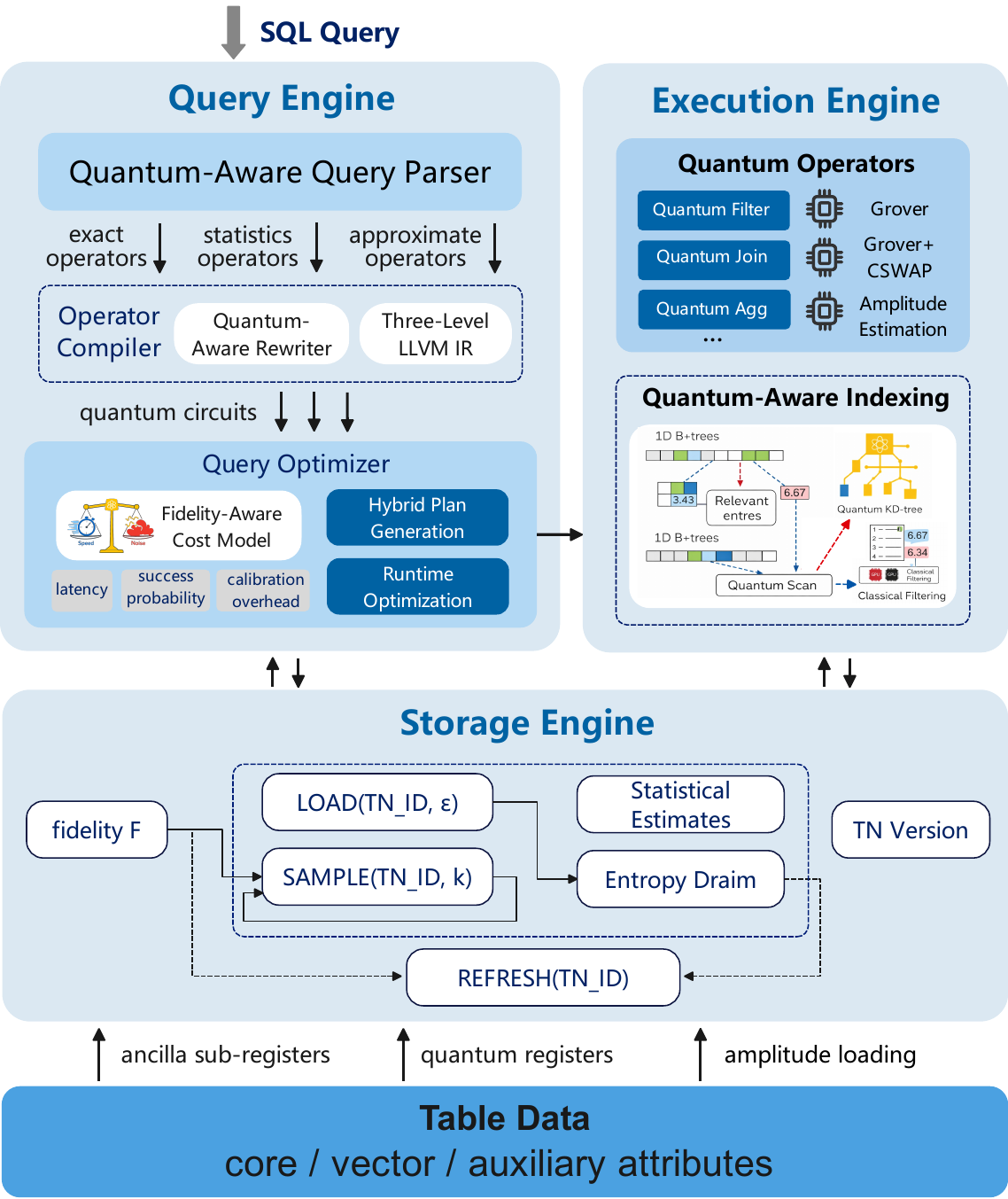}
  \vspace{-.25cm}
  \caption{Example Quantum Database System Design.}
  \label{fig:arch}
  \vspace{-.85cm}
\end{figure}

We present differences between classical and quantum databases (Table~\ref{tbl:compare}) and we outline the key components of \oursys (Figure~\ref{fig:arch}). 

\hi{Design Principles.} Quantum databases follow several design principles that differ from classical ones. (1) Minimize quantum data loading by aggressively shrinking the quantum-active footprint (e.g., via selective indexing, pre-filtering, on-demand encoding). (2) Leverage amplitude-level parallelism instead of tuple-by-tuple processing (e.g., reformulating as quantum operators). (3) Adopt fidelity-aware optimization strategies (e.g., potential noise in quantum operators). (4) Maintain hybrid executability by keeping both quantum and classical paths viable, thereby resolving hardware noise and resource constraints.

\begin{table}[!t]
\caption{Classical vs. Quantum Database Components.}
\vspace{-.2cm}
\label{tbl:compare}
\small
\centering
\resizebox{.48\textwidth}{!}{
\begin{tabular}{p{2cm} p{4.1cm} p{4.1cm}}
\toprule
\multicolumn{1}{c}{\textbf{Component}} & 
\multicolumn{1}{c}{\textbf{Classical Database}} & 
\multicolumn{1}{c}{\textbf{Quantum Database}} \\
\midrule

\textbf{Compiler}
& Generate efficient machine code for classic execution (CPUs, GPUs, ...). 
& Generate circuits executable directly on the quantum backend. \\

\textbf{Operator} 
& Perform tuple-at-a-time or vectorized processing (CPU cost $\downarrow$). 
& Replace tuple scans with amplitude and interference operations. \\ 

\textbf{Plan Optimizer} 
& Select the lowest-cost deterministic execution path. 
& Dynamically balance quantum speedup vs. noise / success rate. \\

\multirow{2}{*}{\textbf{Data Indexing}} 
& \multirow{2}{*}{Accelerate lookup operations.} 
& Shrink the candidate set that must \\

& & be loaded into quantum computer. \\

\multirow{2}{*}{\textbf{Data Format}} 
  & \multirow{2}{*}{Byte-addressable and easy to parse.} 
  & \multirow{2}{*}{Encode data in qubit states.} \\

& & \\

\textbf{Storage Engine} 
& Ensure durability and efficient I/O. 
& Maintain reconstructable large-scale data with compression techniques. \\
\bottomrule
\end{tabular}
}
\vspace{-.55cm}
\end{table}

\hi{Quantum SQL Compiler.} \oursys integrates a quantum-native compilation stack: (1) We handle standard relational operators (e.g., filtering, aggregation) alongside quantum-friendly functions such as sampling-based estimation and probabilistic evaluation, which can be realized using primitives like amplitude estimation~\cite{DBLP:journals/qip/Shyamsundar23}. Note we can automatically generate gate-efficient circuits for each operator via a modular circuit generator~\cite{quanto}; (2) We employ a three-tier LLVM-style intermediate representation (logical IR, quantum-extended IR, and physical circuit IR), supporting stepwise lowering from declarative SQLs to executable quantum circuits~\cite{qir}; (3) We maintain a dynamic catalog of rewrite rules that uncovers useful transformations for complex joins, and multi-predicate filter queries via specific quantum features like superposition-enabled probabilistic branching and interference-based correlation detection.

\hi{Quantum Plan Optimizer.} \oursys models each quantum operator $q$ using a compact profile $(T_q, P_q, \varepsilon_q)$ that summarizes its latency, success probability, and approximation error, allowing the optimizer to compare it directly with classical operators. Using these profiles, execution plans embed both quantum and classical realizations of eligible operators, thereby keeping domain choices flexible. At runtime, \oursys monitors quality and stability, adapting to noise by increasing shots, switching to more robust circuit variants, or falling back to classical paths when quantum performance degrades.

\hi{Quantum-Accelerable Operators.} \oursys offloads costly operations to quantum subroutines that encode computation into probability amplitudes rather than row-wise execution~\cite{quantumANN, SWAPTEST}. For instance, for similarity joins, \oursys bypasses dimension-wise traversal by encoding vectors into quantum states and estimating their inner products through interference-based overlap.




\hi{Quantum-Aware Indexing.} To address the quantum memory bottleneck, we develop index structures that enable \emph{selective quantum probing}, meaning that only index entries relevant to the query are probed before allocating quantum resources. We begin with lightweight quantum scans over one-dimensional B$^+$-trees to quickly narrow the candidate set. When this yields only a small number of candidates, the remaining work is completed via classical filtering. However, if many candidates persist, we transition to a quantum KD-tree that jointly indexes multiple dimensions, enabling more aggressive pruning without exhaustively scanning all rows.

\hi{Quantum Data Format.} \textit{(1) Basis Encoding:} Core row identifiers like $ID_{row}$ are mapped to computational basis states, enabling uniform superposition and parallel oracle processing. \textit{(2) Amplitude Encoding:} Data vectors (e.g., embeddings) are encoded into state amplitudes via rotation gates, compressing values into probability distributions aligned with row indices. \textit{(3) Control Encoding:} Metadata and flags are stored in ancilla qubits that act as control lines, enabling conditional gate execution based on row-level attributes.


\hi{Quantum Storage Engine.} Unlike page-based designs~\cite{quantable}, \oursys adopts a state-centric approach. Quantum-active attributes are stored as compressed tensor networks (TNs), acting as logical pages that support bounded reconstruction and sampling under fidelity constraints. The engine provides fidelity-aware primitives to enable controlled approximation while tracking entropy and noise: \texttt{LOAD}(TN\_ID, $\epsilon$) for approximate access, \texttt{SAMPLE}(TN\_ID, $k$) for bounded sampling, and \texttt{REFRESH}(TN\_ID) to restore fidelity when degradation exceeds threshold. Classical metadata is stored separately to ensure recoverability.

\section{Quantum-Accelerable Operators}
\label{sec:operations}

Building on the basic circuits introduced in Section~\ref{sec:quantum_basics}, we design quantum circuits  tailored to core database operations. Each operator is restructured to exploit quantum parallelism while preserving relational semantics. Table~\ref{tab:quantum_ops_cost} summarizes the quantum-accelerable operators and we illustrate three representative types below.

\begin{table*}[!t]
\vspace{-1.cm}
\raggedright
\caption{Quantum-Accelerable Relational Operators ($M$ and $N$ are total and qualifying row numbers; $d$ is the vector dimension; $\epsilon$ is precision; $n$ and $q_{*}$  are qubit counts for data and precision registers; $a_{*}$ is ancilla; $b$ is bandwidth; $D$ is the gate depth).}
\vspace{-.3cm}
\label{tab:quantum_ops_cost}
\footnotesize

\renewcommand{\arraystretch}{0.78}
\setlength{\tabcolsep}{2.3pt}
\setlength{\extrarowheight}{0pt}

\renewcommand{\tabularxcolumn}[1]{>{\centering\arraybackslash}m{#1}}
\newcolumntype{Y}{X}

\begin{tabularx}{\textwidth}{%
>{\raggedright\arraybackslash}m{1.1cm} 
>{\centering\arraybackslash}m{1.5cm}   
>{\centering\arraybackslash}m{2.6cm}   
>{\centering\arraybackslash}m{2.7cm}   
>{\centering\arraybackslash}m{1.4cm}   
>{\centering\arraybackslash}m{1.8cm}   
>{\centering\arraybackslash}m{1.7cm}   
Y                                      
}
\toprule
\multirow{2}{*}{\thead{Type}} &
\multirow{2}{*}{\thead{Operator\\(SQL)}} &
\multirow{2}{*}{\thead{Typical Scenario}} &
\multirow{2}{*}{\thead{Candidate\\Quantum Algorithm}} &
\multicolumn{2}{c}{\makecell[c]{\textbf{Complexity}}} &
\multicolumn{2}{c}{\makecell[c]{\textbf{Cost Model}}} \\
\cmidrule(lr){5-6} \cmidrule(l){7-8}

& & & & 
\makecell[c]{\textbf{Quantum}} & 
\makecell[c]{\textbf{Classical} \\\textbf{(without index)}} &
\makecell[c]{\textbf{Qubit Num}} &
\makecell[c]{\textbf{Gate Depth}\\\textbf{(per-circuit call)}} \\
\midrule

\multirow{4}{*}[-1em]{\shortstack[l]{\textbf{Filter}}} & Equality Filter &
\texttt{WHERE column = value} &
\makecell[c]{Grover~\cite{Grover} (Search)} &
$O(\sqrt{N})$ &
$O(N)$ &
$n + a_{\mathrm{orc}}$ &
\makecell[c]{$O\!\left((D_{\mathrm{orc}} + D_{\mathrm{diff}})\cdot \sqrt{N/M}\right)$} \\

& Range Filter  
&
\texttt{WHERE column > value} &
\makecell[c]{Grover~\cite{Grover} \\ (Threshold Oracle)} &
$O(\sqrt{N})$ &
$O(N)$ &
$n + b + a_{\mathrm{cmp}}$ &
\makecell[c]{$O\!\left((D_{\mathrm{cmp}}(b) + D_{\mathrm{diff}})\cdot \sqrt{N/M}\right)$} \\

& LIKE &
\begin{tabular}[t]{@{}c@{}}\texttt{WHERE column LIKE `val\%'}\end{tabular} & \makecell[c]{
Grover~\cite{Grover}\\(Prefix-Match Oracle)} &
$O(\sqrt{N})$ &
$O(N)$ &
$n + b + a_{\mathrm{pref}}$ &
\makecell[c]{$O\!\left((D_{\mathrm{pref}}(b) + D_{\mathrm{diff}})\cdot \sqrt{N/M}\right)$} \\

& EXISTS & Existence check &
\makecell[c]{Grover~\cite{Grover}\\(Quantum Counting)} &
$O(\sqrt{N})$ &
$O(N)$ &
\makecell[c]{$n + a_{\mathrm{orc}}$} &
\makecell[c]{$O\!\left((D_{\mathrm{orc}} + D_{\mathrm{diff}})\cdot \sqrt{N}\right)$} \\
\cmidrule(l){1-8}

\multirow{2}{*}[-0.95em]{\thead{Join}} & Equi-join & \begin{tabular}[t]{@{}c@{}}\texttt{\texttt{column1 = column2}}\end{tabular} &
\makecell[c]{Grover~\cite{Grover}\\(Index Probing)} &
\makecell[c]{$O(\sqrt{N_2})$ / probe} &
$O(N_1 \cdot N_2)$ &
$n_2 + b_{\mathrm{key}} + a_{\mathrm{orc}}$ &
\makecell[c]{$O\!\left((D_{\mathrm{idx}} + D_{\mathrm{key}} + D_{\mathrm{diff}}(n_2))\cdot \sqrt{N_2}\right)$\\
/probe} \\

& Non-equi join & \begin{tabular}[t]{@{}c@{}}\texttt{column1} $\leq$ \texttt{column2}\end{tabular} &
\makecell[c]{Grover~\cite{Grover}\\(Comparison Oracle)} &
\makecell[c]{O($\sqrt{N_2}$) / probe} &
$O(N_1 \cdot N_2)$ &
$n_2 + b_{\mathrm{key}} + b + a_{\mathrm{cmp}}$ &
\makecell[c]{$O\!\left((D_{\mathrm{cmp}} + D_{\mathrm{diff}}(n_2))\cdot \sqrt{N_2}\right)$\\
/probe} \\
& Similarity Join &
Similarity search &
\makecell[c]{Grover~\cite{Grover}\\(SWAP Test)} &
$O(\sqrt{N})$ &
$O(d \cdot N)$ &
$2q + a_{\mathrm{prep}} + 1$ &
\makecell[c]{$O\!\left(2\cdot D_{\mathrm{prep}}(d) + O(q)\right)$} \\
\cmidrule(l){1-8}

\multirow{5}{*}[-1.5em]{\thead{Aggregation}} & MIN &
Global minimum search &
\makecell[c]{Dürr--Høyer\\Minimum Finding~\cite{durr1996quantum}} &
$O(\sqrt{N})$ &
$O(N)$ &
$n + b + a_{\mathrm{cmp}}$ &
\makecell[c]{$O\!\left((D_{\mathrm{cmp}} + D_{\mathrm{diff}})\cdot \sqrt{N}\right)$} \\

& COUNT &
Cardinality estimation &
Amplitude Estimation~\cite{AE} &
$O(1/\epsilon)$ &
$O(N)$ &
$n + a_{\mathrm{orc}} + q_{\epsilon}$ &
\makecell[c]{$O\!\left((D_{\mathrm{orc}} + D_{\mathrm{diff}})\cdot (1/\epsilon)\right)$} \\ \\

& SUM &
Total value calculation&
Normalization + Amplitude Estimation~\cite{AE} &
$O(1/\epsilon)$ &
$O(N)$ &
$n + b_v + a_{\mathrm{rot}} + q_{\epsilon}$ &
\makecell[c]{$O\!\left((D_{\mathrm{load}} + D_{\mathrm{rot}} + D_{\mathrm{diff}})\cdot (1/\epsilon)\right)$} \\

& AVG &
Mean value calculation&
Normalization + Amplitude Estimation~\cite{AE} &
$O(1/\epsilon)$ &
$O(N)$ &
$n + b_v + a_{\mathrm{rot}} + q_{\epsilon}$ &
\makecell[c]{$O\!\left((D_{\mathrm{load}} + D_{\mathrm{rot}} + D_{\mathrm{diff}})\cdot (1/\epsilon)\right)$} \\

\cmidrule(l){1-8}

\multirow{1}{*}[-0.em]{\thead{Other}}& Sampling &
Approximate query &
Amplitude Estimation~\cite{AE} &
$O(\sqrt{N})$ &
$O(N)$ &
$n + a_{\mathrm{orc}}$ &
\makecell[c]{$O\!\left((D_{\mathrm{orc}} + D_{\mathrm{diff}})\cdot \sqrt{N/M}\right)$} \\

\bottomrule
\end{tabularx}
\vspace{-.25cm}
\end{table*}

\subsection{Quantum Data Filtering}
\label{subsec:scan_filter}

Large-scale data filtering is a basic database operation, yet it incurs high latency on classical systems when processing sparse, unindexed data because of exhaustive full-table scans. To overcome this bottleneck, \oursys adapts Grover’s algorithm to a quantum data filtering operator.
We design a custom quantum circuit that implements a sequence of logic and diffusion gates, replacing the classical linear scan with an accelerated quantum search.


\hi{Circuit Design.} Specifically, we map the row identifier ($ID_{row}$) domain as the search space and compile $\phi$ into a quantum oracle $O_\phi$ that marks satisfying rows (as $O_f$ in Section~\ref{sec:quantum_basics}). Compound SQL predicates are transformed into phase-marking logic via ancilla-assisted encoding. For example, a predicate such as \texttt{age > 30 AND city = `NY'} is implemented by first evaluating each subcondition into ancilla qubits and then combining them using multi-controlled operations to mark the final target states. After Grover iterations amplify the amplitude of matched $ID_{row}$, \oursys performs classical reconciliation, i.e., deduplicating matched $ID_{row}$ and rechecking them against $\phi$ to guarantee correctness.

\subsection{Quantum Similarity Join}
\label{subsec:join_match}



Similarity joins are fundamental in databases and multi-modal data retrieval~\cite{DiskJoin, simjoin, simjoin2, simjoin3}. They aim to retrieve all pairs of vectors from two datasets whose distances are within a given threshold. The core operation of a similarity join is distance computation (e.g., inner product), which determines the set of vectors that are similar (i.e., with small distance) to each input vector. 
Classical machines typically incur high costs to compute inner products between vectors (e.g., due to repeated dimension-wise multiplication and distance evaluations across numerous vector pairs). Instead, \oursys encodes vectors into quantum states and computes the inner product without explicitly traversing all dimensions or candidate vector pairs.

\begin{figure}[!t]
  \vspace{-1.cm}
  \centering
  \includegraphics[width=0.95\linewidth]{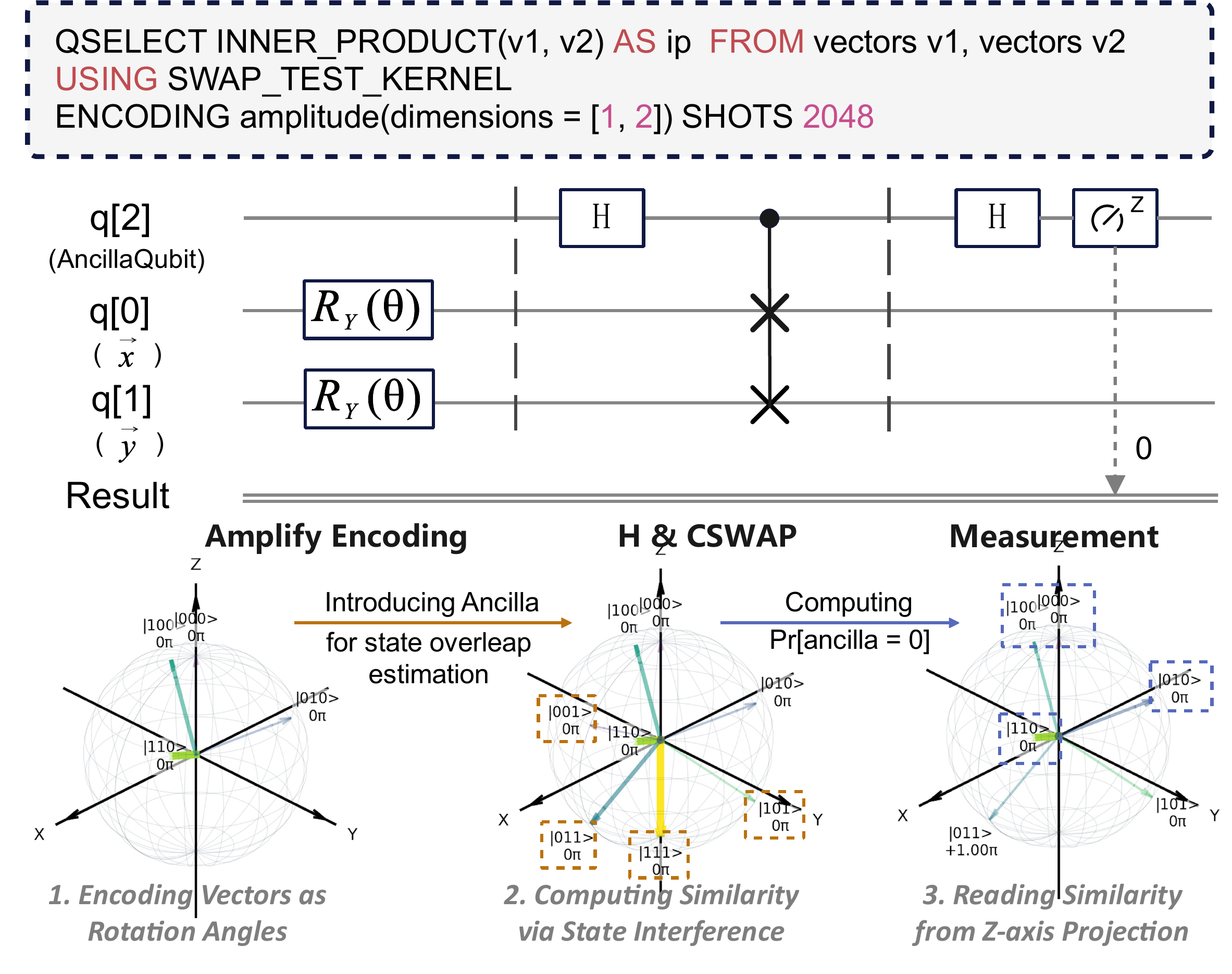}
  \vspace{-.35cm}
  \caption{CSWAP-Gate Based Inner Product Calculation.}
  \Description{\sys Inner Product}
  \label{fig:swapInnerproduct}
  \vspace{-.45cm}
\end{figure}

\hi{Circuit Design.} As illustrated in Figure~\ref{fig:swapInnerproduct}, given two normalized vectors $\vec{x}, \vec{y} \in \mathbb{R}^d$, \oursys first encodes them into quantum states $\ket{x}$ and $\ket{y}$. Each vector is transformed into a quantum state using parameterized single-qubit rotations (e.g., $R_Y(\theta)$), following the vector-to-state mapping described in Section~\ref{sec:overview}. This encoding ensures that the inner product between $\vec{x}$ and $\vec{y}$ is reflected in the overlap between the corresponding quantum states. To estimate this overlap, \oursys employs the \emph{SWAP Test}~\cite{SWAPTEST}, a standard quantum subroutine for inner product estimation. Specifically, an ancilla qubit is initialized into a uniform superposition using an H gate, after which a CSWAP operation is applied between the two data registers conditioned on the ancilla. A second H gate is then applied to the ancilla qubit. Upon measuring the ancilla qubit, the probability of observing outcome $\ket{0}$ is given by
$\Pr[\text{ancilla}=0] = (1 + |\langle x | y \rangle|^2)/2$.
By repeatedly executing the circuit and measuring the ancilla qubit, \oursys estimates $|\langle x | y \rangle|^2$ from the empirical measurement frequency. Classical post-processing converts measurement outcomes into approximate inner products for identifying vector pairs.

\subsection{Quantum Aggregation}
\label{subsec:agg_opt}


\begin{sloppypar}
Classical aggregations (e.g., \texttt{SUM}) compute results by iterating through all qualifying rows and accumulating their values.
It incurs high latency when the qualifying set is large, as it scales linearly with the row volumes.
Quantum systems encode information as probability amplitudes, making it infeasible to replicate a classical accumulator directly. 
To address this, \oursys~reformulates aggregation as a probability estimation task {so that I/O does not dominate end-to-end cost with coherent vector access (e.g. via QRAM)}. By mapping numeric values to the observation probability of a target qubit, it utilizes amplitude estimation~\cite{AE} to compute the global sum efficiently without iterating through individual rows.
\end{sloppypar}

\hi{Circuit Design.} \oursys first prepares a uniform superposition over all $N$ candidate row identifiers. For each row identifier $\ket{x}$, the corresponding value $v(x)$ is accessed and used to control a dedicated flag qubit, referred to as the \emph{Good} qubit. Specifically, a controlled rotation is applied so that the probability of the Good qubit being measured as $\ket{1}$ is proportional to the normalized value $v(x) / V_{\max}$, where $V_{\max}$ is a known upper bound on the aggregated values. Intuitively, rows with larger values contribute more strongly to the probability of observing the Good flag. The overall probability of measuring the Good qubit as $\ket{1}$ corresponds to the average of the normalized values across all rows, i.e., $Pr[{\mathrm{Good} = 1]} = \frac{1}{N} \sum_{x=0}^{N-1} \frac{v(x)}{V_{\max}}$. By estimating using amplitude estimation and scaling it by $N$ and $V_{\max}$, \oursys recovers the desired result:
$\mathrm{SUM}(v) = N \cdot V_{\max} \cdot \Pr[\text{Good}=1]$.

\begin{sloppypar}
    
\section{Quantum-Aware Index Structure}
\label{sec:qindex}

Due to limited qubit capacity and circuit depth, existing quantum computers can encode only a small fraction of data in superposition~\cite{qudatadiscuss}, making selective data loading essential. 
Existing approaches load only relevant B$^+$-tree leaf nodes into the quantum processor, but this strategy generalizes poorly to multi-dimensional queries and suffers significant performance degradation.
We therefore propose a quantum-accelerated indexing strategy that narrows the query scope early via lightweight probing, supporting both one- and multi-dimensional workloads.

\begin{figure}[!t]
    \vspace{-1.25cm}
    \centering
    \includegraphics[width=\linewidth, trim=0 0 0 0, clip]{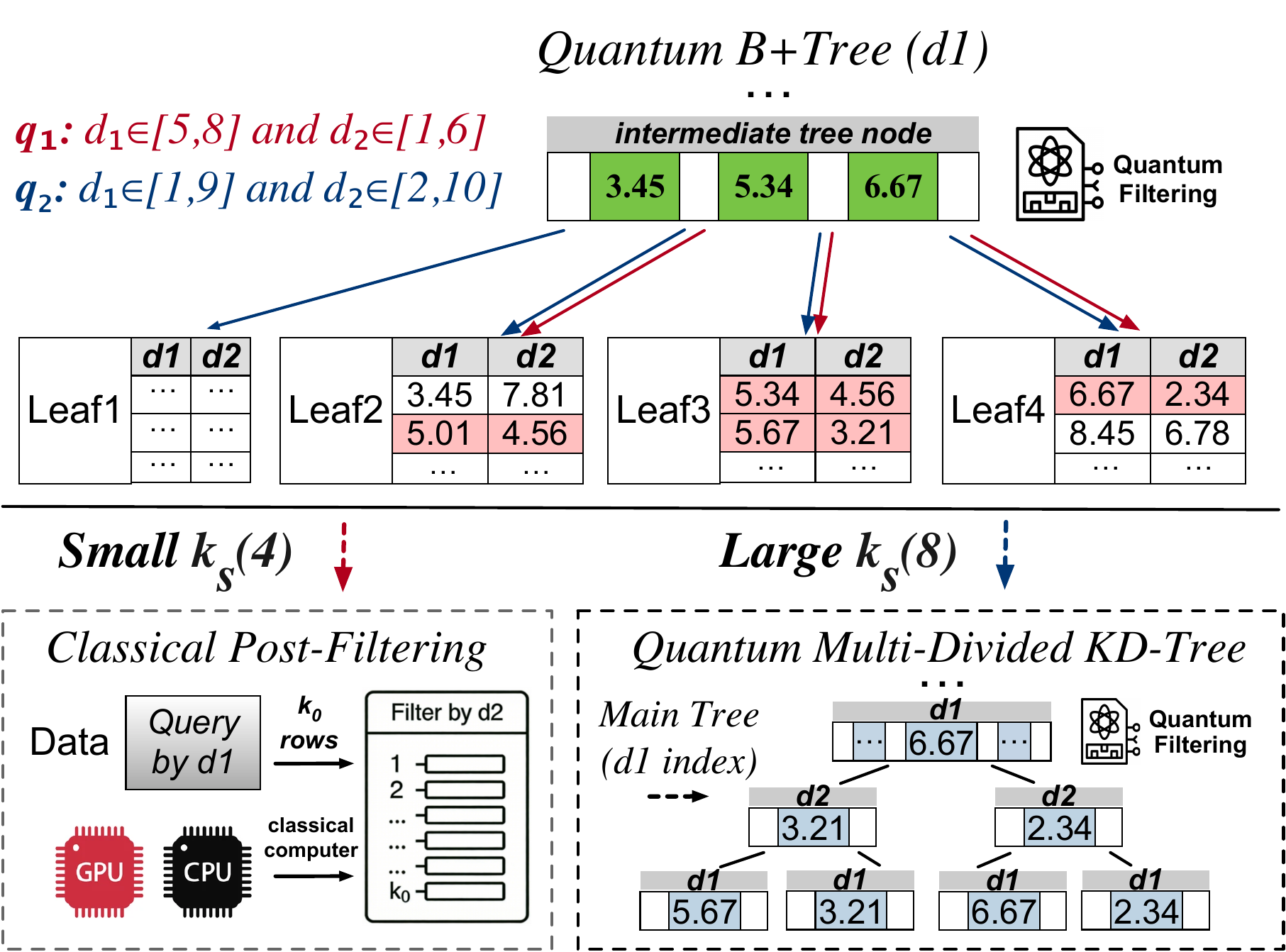}
    \vspace{-.5cm}
    \caption{Quantum-Aware Multi-Dimensional Index.}
    
    \label{fig:Qb+tree}
    \vspace{-1cm}
\end{figure}

Our strategy integrates quantum probing on selected dimensions with classical post-filtering. For multi-column predicates (Figure~\ref{fig:Qb+tree}), we first probe several one-dimensional quantum B$^+$-trees constructed on selected dimensions. This probing identifies nodes that are fully contained in, or partially overlap with, the query range~\cite{quantum-bptree-2024}. For example, for the query $q_1: d_1 \in [5, 8]$, a quantum probe returns $k_1 = 4$ tuples satisfying the predicate on $d_1$. Among the result sets returned by all probed B$^+$-trees, we select the smallest one as the candidate set, and denote its size by $k_s$. Based on $k_s$, we dynamically choose between two strategies.

\noindent $\bullet$ \textbf{(1) Classical Post-Filtering.} If the number of candidates is small (i.e., \(k_s = O(\log N)\)), we perform classical verification on the remaining \(d - 1\) dimensions. For query \(q_1\), we further filter the 4 retrieved rows using the secondary predicate \(d_2 \in [1, 6]\). It incurs a cost of \(O(k_s \cdot (d - 1))\), efficient due to the small candidate size. 

\noindent $\bullet$ \textbf{(2) Quantum Multi-Divided KD-Tree Search.} If \(k_s\) is large, classical filtering becomes less effective. We escalate to a quantum multi-divided KD-tree that jointly indexes multiple dimensions. 
By recursively partitioning the data along different dimensions and applying quantum-assisted probing at each level, this structure significantly reduces the candidate set before verification, making it more suitable for large \(k_s\) scenarios.

Additionally, for queries involving disjunctive conditions (e.g., \(d_1 \in [5,8] \lor d_2 \in [2,5]\)), we independently probe each dimension and combine the resulting superpositions. 

\end{sloppypar}

\section{Quantum-Aware Optimizer}
\label{sec:optimizer}

Classical query optimizers rely on deterministic assumptions that fail to capture the stochasticity and noise inherent in quantum hardware. \oursys~addresses this limitation by treating quantum operators as probabilistic, accuracy-bounded primitives, thereby shifting the optimization goal from pure latency minimization to a robust trade-off between speed and reliability. 




\hi{Estimation Model.} The cost model extends classical optimization by treating a quantum operator as a stochastic, accuracy‑bounded primitive rather than a deterministic function. The optimizer evaluates each operator using a compact tuple $(T_q, P_q, \varepsilon_q)$ that captures its latency, success probability, and approximation error, 
which are quantified directly from circuit structure and hardware characteristics. Each factor is explicitly incorporated into the objective function, allowing the optimizer to trade off execution speed against expected reliability under real hardware behavior.

\noindent$\bullet$ \textbf{(1) Time Estimation ($\boldsymbol{T_q}$).} Execution time is derived from the circuit’s layerized schedule after topology‑aware routing. For layers $L=\{L_1,\ldots,L_K\}$, \(T_q = \sum_{k=1}^{K} \Bigl( \max_{g \in L_k} t_g + t_{\text{ctrl}} \Bigr)\), where $t_g$ is the calibrated duration of each gate and $t_{\text{ctrl}}$ accounts for control and synchronization overheads between layers.
It involves two latency contributors: (i) serialization imposed by limited qubit connectivity and (ii) SWAP-induced depth inflation, both of which are used to assess whether a quantum operator outperforms the classical ones.

\noindent$\bullet$ \textbf{(2) Noise and Success Probability ($\boldsymbol{P_q}$).} Quantum success probability is estimated as a circuit-level composition of gate errors and decoherence. For each layer, \(p_k = \Bigl( 1 - \sum_{g \in L_k} \epsilon_g \Bigr)\cdot
      \exp\Bigl(- \frac{t_k}{T_2^{\text{eff}}}\Bigr)\), with $\epsilon_g$ the hardware‑reported gate error rates and $T_2^{\text{eff}}$ an effective coherence bound. The operator-level success probability is approximated by $P_q = \prod_{k=1}^{K} p_k$. It ties plan quality directly to circuit depth and layout: deeper or poorly routed circuits incur multiplicative degradation, guiding the optimizer to prune overly fragile plans.

\noindent$\bullet$ \textbf{(3) Hybrid Execution Error ($\boldsymbol{\varepsilon_q}$).} Each quantum operator is paired with a deterministic classical fallback. The optimizer evaluates its expected runtime as \(\mathbb{E}[T_q] = 
P_q \cdot T_q^{\text{quantum}}
+ (1 - P_q) \cdot T_q^{\text{classical}}
\), ensuring that incorrect or failed executions do not compromise result correctness. The approximation error $\varepsilon_q$, derived from the operator’s amplitude, estimation precision, or sampling budget, is incorporated as a constraint or penalty depending on the query’s accuracy requirements. Parameters are updated using runtime observations, enabling the optimizer to refine plan choices adaptively.

By grounding each component of $(T_q, P_q, \varepsilon_q)$ in explicit circuit and hardware features, the model provides a concrete and actionable basis for quantum-classical plan cost modeling.

\hi{Hybrid Plan Generation.} In \oursys, each plan encodes both quantum and classical implementations of selected operators, preserving semantic equivalence while exposing explicit alternatives for cost and quality trade-offs. Importantly, cross-domain coordination costs (e.g., data movement between classical and quantum runtimes) are explicitly incorporated into the optimizer's objective function, preventing overly optimistic cost estimates. Furthermore, \oursys supports deferred binding of execution-domain choices, allowing the runtime dispatcher to select the most suitable realization based on real-time conditions (e.g., backend calibration state, queue delays).

\hi{Adaptive Runtime Optimization.} To ensure robust execution in the presence of quantum noise and stochastic failures, \oursys integrates adaptive runtime optimization as a first-class mechanism. When a quantum path fails to meet declared quality or latency thresholds, the system dynamically reallocates effort by increasing the number of sampling shots, switching to more robust circuit variants, or falling back to classical computation. Critically, all such adaptations operate under bounded-error semantics: fallback paths are validated for correctness in advance, and all runtime decisions are constrained by the optimizer's declared error and latency budgets. This architecture ensures that query results remain trustworthy even when quantum hardware fails unpredictably, allowing \oursys to balance optimism with reliability in real deployments.

\section{Preliminary Results}

\hi{Prototype Implementation.}
We implement a minimal prototype of \oursys that supports database filtering operations in~\cite {QuteLink}. The classical execution engine handles SQL parsing, query planning, and result reconciliation, while data filtering is offloaded to a quantum computing backend. The prototype is implemented using the \texttt{QPanda3} library (v1.0) and developed on the \texttt{origin\_wukong} platform. All quantum circuits are executed on a real QPU with 72 qubits and noise. Each quantum experiment is executed with 2000 measurement shots by default, and the results are averaged.


\hi{Dataset.} 
To evaluate the effectiveness of quantum computing over large-scale data, we construct a synthetic dataset with $N = 2^{40}$ tuples using a random generation method. We consider filter queries of the form \texttt{SELECT RID FROM T WHERE $\phi(x)$}, where $\phi$ consists of single-attribute or conjunctive predicates over the generated attributes. Predicate parameters are chosen to control query selectivity, and the selectivity of each predicate is bounded by 2\%, with multiple selectivity levels generated within this bound.

\hi{Evaluation Methods.}
{Given that classical database runtime is highly sensitive to hardware and system optimizations, we use a cost-model-based evaluation for a fair comparison of algorithmic complexity.
Specifically, we evaluate two approaches: (1) a classical database baseline, whose performance is estimated using a traditional cost model~\cite{vldb02}; and (2) \oursys, whose performance is measured using actual execution when the data size permits and otherwise estimated using our cost model (Section~\ref{sec:optimizer}).}

\begin{figure}[!t]
	\centering
	\includegraphics[width=.95\linewidth]{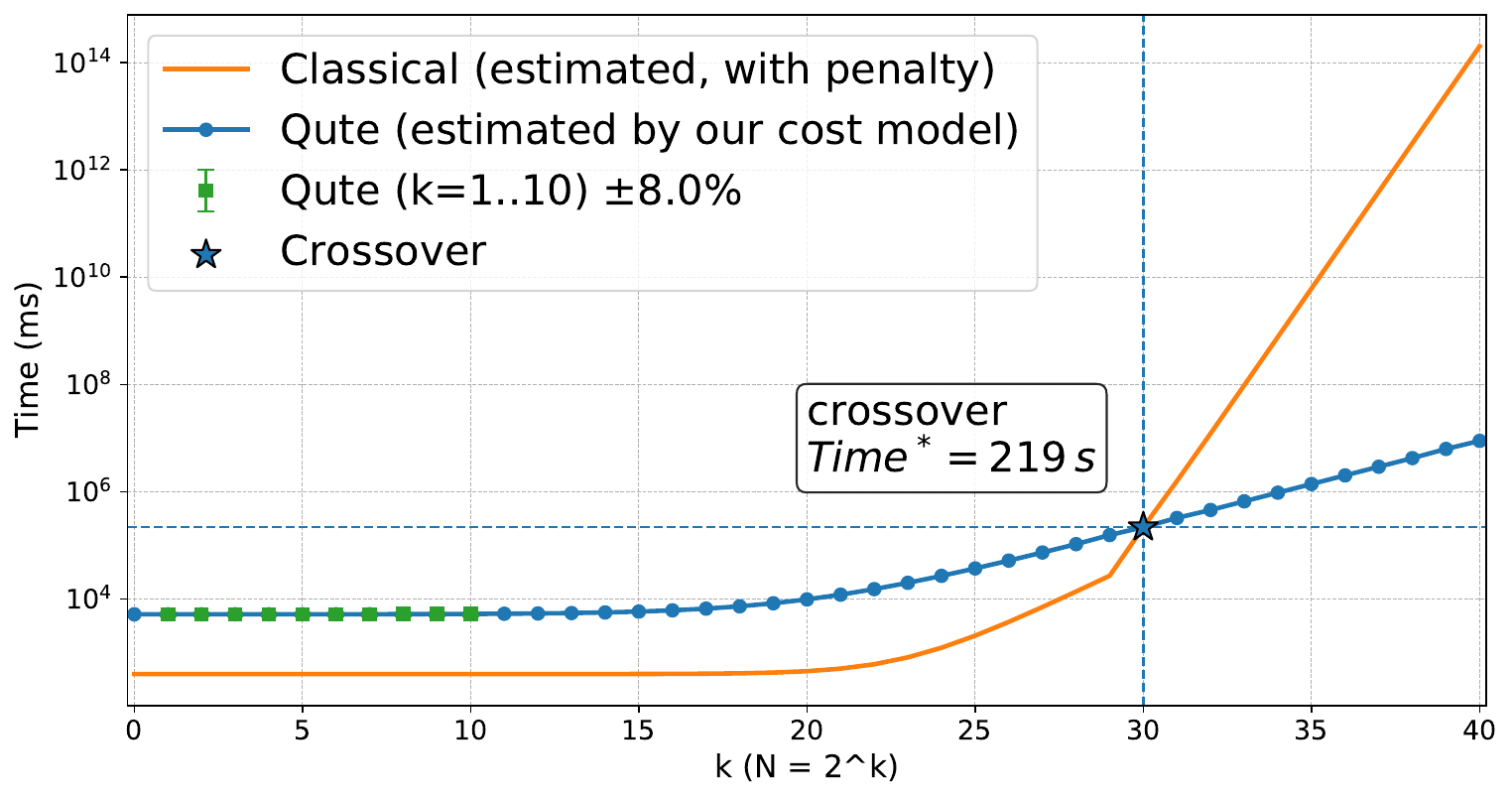}
	\vspace{-.25cm}
	\caption{Performance Comparison (\oursys vs. Classic). 
    }
	\label{fig:exp}
	\vspace{-.55cm}
\end{figure}

\hi{Results.} 
Figure~\ref{fig:exp} compares the processing time of \oursys and a classical database under increasing data scale. 
For \oursys, the measured runtime on small datasets (up to $2^{10}$) is shown as discrete sample points with a Grover error of $\pm 8.0\%$, while the runtime estimated by our cost model is shown as a continuous curve. 
The close agreement between the measured and estimated results indicates that the cost model accurately captures the performance of \oursys and can be reliably used for large-scale extrapolation. 
The estimated performance of the classical database is also reported for comparison.
The results reveal a clear crossover point at approximately $N \approx 2^{30}$. 
For smaller datasets, the classical database is faster due to lower constant overhead, whereas for larger datasets, \oursys consistently outperforms the classical approach. 
It highlights the scalability advantage of \oursys for extreme-scale, unindexed filtering workloads.

\section{Evolution and Further Directions}
\label{sec:discussion}


We classify the evolution of quantum databases into three stages (S1-S3) based on data scale and computational capability (Figure~\ref{fig:intro}).

\noindent \ding{182} \hi{Quantum-Assisted Database.} Quantum computing functions as a co-processor to classical systems. Due to limited qubits, short coherence times, and high data-loading costs, quantum resources are used only for selective, compute-intensive tasks such as full-table scans using Grover's search. Data remains in classical memory, with quantum execution triggered in real time. There is no index acceleration, and fallback mechanisms ensure correctness when quantum runs fail or exceed noise thresholds. The primary bottleneck lies in the classical-to-quantum transfer bandwidth.


\noindent \ding{183} \hi{Quantum-Centric Database.}
With improved quantum memory and loading methods, quantum computing becomes the main execution engine. Hybrid quantum-classical indexes emerge, enabling selective access to superposition states. A fidelity-aware optimizer plans across quantum and classical paths using cost models that account for decoherence, gate latency, and qubit limits. Routing mechanisms dispatch subqueries accordingly. Core operations such as joins and aggregations are implemented using quantum primitives, including SWAP tests and Amplitude Estimation, enabling efficient hybrid execution.


\noindent \ding{184} \hi{Quantum-Native Database.} At this stage, fault-tolerant quantum hardware supports more powerful storage and computation. All data types (including structured, unstructured, multimodal) reside in quantum memory. Advanced data encodings (e.g., vector quantization algorithms) compress and manipulate large datasets directly. Classical components are minimized, and the system achieves fully quantum-native storage, indexing, and processing.

\section{Related Work}
\label{sec:related}

\noindent\textbf{Quantum Computation for Databases.} Prior works include surveys and efforts in query processing~\cite{DBLP:conf/icde/0003HF24,DBLP:journals/pvldb/CalikyilmazGGWPSAPG23}, query optimization~\cite{queryoptimization1,queryoptimization2,queryoptimization3,queryoptimization4,queryoptimization5,liu2025q2o}, index selection~\cite{indexselection1,indexselection2}, data integration~\cite{dataintegration1}, and transaction scheduling~\cite{transactionscheduling1,transactionscheduling2,transactionscheduling3}. Other efforts explore quantum-assisted approximate query processing~\cite{quanmeetdb} and data structures~\cite{quandatastructure}. While promising, they largely retain classical data models and storage layers, using quantum computation as an isolated acceleration step without cross-layer integration.

\noindent\textbf{Databases for Quantum Computation.} Several works explore how classical database abstractions can support quantum computation. This includes storing and manipulating relational data on quantum hardware~\cite{rieger2024operationalqdb,quantable}, compiling SQL queries to quantum circuits~\cite{quanviable}. Quantum machine learning has also been applied to relational and graph data~\cite{QML1,QML2,QML3,QML4}, though these focus on model training rather than database execution. 

\section{Conclusion}
\label{sec:conclusion}


This paper envisions quantum-native databases as a foundational shift in data management, in which quantum computation is integrated across the full stack, from query parsing to execution. We proposed \oursys, a system that addresses key challenges in operator compilation, hybrid optimization, selective indexing, and fidelity-aware storage. Unlike prior simulation-based approaches, \oursys demonstrates a unified design that supports practical quantum query execution. We outlined a three-stage roadmap highlighting both architectural milestones and research opportunities for the future of quantum data systems.



\clearpage
\balance
\bibliographystyle{ACM-Reference-Format}
\bibliography{refs}

@inproceedings{quantable,
  author       = {Tuodu Li and
                  Gongsheng Yuan and
                  Chang Yao and
                  Meng Shi and
                  Ziyue Wang and
                  Ling Qian and
                  Jiaheng Lu},
  title        = {Quantum Storage Design for Tables in {RDBMS}},
  booktitle    = {{VLDB} Workshops},
  publisher    = {VLDB.org},
  year         = {2024}
}

@article{quandatastructure,
  title={Towards Quantum Data Structures for Enhanced Database Performance},
  author={Littau, Tim and Li, Ziyu and Hai, Rihan},
  journal={VLDB 2024 Workshop},
  volume={2150},
  pages={8097}
}

@article{qudatadiscuss,
  author       = {Rihan Hai and
                  Shih{-}Han Hung and
                  Tim Coopmans and
                  Tim Littau and
                  Floris Geerts},
  title        = {Quantum Data Management in the {NISQ} Era},
  journal      = {Proc. {VLDB} Endow.},
  volume       = {18},
  number       = {6},
  pages        = {1720--1729},
  year         = {2025}
}

@article{quandbsurvey,
  author       = {Gongsheng Yuan and
                  Yuxing Chen and
                  Jiaheng Lu and
                  Sai Wu and
                  Zhiwei Ye and
                  Ling Qian and
                  Gang Chen},
  title        = {Quantum Computing for Databases: Overview and Challenges},
  journal      = {CoRR},
  volume       = {abs/2405.12511},
  year         = {2024}
}

@inproceedings{DBLP:conf/icde/0003HF24,
  author       = {Rihan Hai and
                  Shih{-}Han Hung and
                  Sebastian Feld},
  title        = {Quantum Data Management: From Theory to Opportunities},
  booktitle    = {{ICDE}},
  pages        = {5376--5381},
  publisher    = {{IEEE}},
  year         = {2024}
}

@inproceedings{quanviable,
  author       = {Manish Kesarwani and
                  Jayant R. Haritsa},
  title        = {Is Quantum-Based {SQL} Query Execution Viable?},
  booktitle    = {{VLDB} Workshops},
  publisher    = {VLDB.org},
  year         = {2024}
}

@article{quanmeetdb,
  author       = {Sai Wu and
                  Meng Shi and
                  Dongxiang Zhang and
                  Junbo Zhao and
                  Gongsheng Yuan and
                  Gang Chen},
  title        = {When Quantum Computing Meets Database: {A} Hybrid Sampling Framework
                  for Approximate Query Processing},
  journal      = {{IEEE} Trans. Knowl. Data Eng.},
  volume       = {36},
  number       = {12},
  pages        = {9532--9546},
  year         = {2024}
}

@article{liu2025q2o,
  title   = {A Demonstration of Q\textsuperscript{2}O: Quantum‑augmented Query Optimizer},
  author  = {Liu, Hanwen and Spedalieri, Federico M. and Sabek, Ibrahim},
  journal = {Proceedings of the VLDB Endowment},
  volume  = {18},
  number  = {12},
  pages   = {5439--5443},
  year    = {2025},
  doi     = {10.14778/3750601.3750691}
}

@article{DBLP:journals/pvldb/CalikyilmazGGWPSAPG23,
  author       = {Umut {\c{C}}alikyilmaz and
                  Sven Groppe and
                  Jinghua Groppe and
                  Tobias Winker and
                  Stefan Prestel and
                  Farida Shagieva and
                  Daanish Arya and
                  Florian Preis and
                  Le Gruenwald},
  title        = {Opportunities for Quantum Acceleration of Databases: Optimization
                  of Queries and Transaction Schedules},
  journal      = {Proc. {VLDB} Endow.},
  volume       = {16},
  number       = {9},
  pages        = {2344--2353},
  year         = {2023}
}

@article{queryoptimization1,
  author    = {Immanuel Trummer and Christoph Koch},
  title     = {Multiple Query Optimization on the {D-Wave} 2{X} Adiabatic Quantum Computer},
  journal   = {Proceedings of the VLDB Endowment},
  volume    = {9},
  number    = {9},
  pages     = {684--695},
  year      = {2016},
  doi       = {10.14778/2947618.2947622}
}

@article{queryoptimization2,
  author    = {Tobias Fankhauser and Marc E. Sol{\`e}r and Rudolf M. F{\"u}chslin and Kurt Stockinger},
  title     = {Multiple Query Optimization Using a Hybrid Approach of Classical and Quantum Computing},
  journal   = {arXiv preprint arXiv:2107.10508},
  year      = {2021},
  url       = {https://arxiv.org/abs/2107.10508}
}

@article{queryoptimization3,
  author    = {Tobias Fankhauser and Marc E. Soler and Rudolf M. F{\"u}chslin and Kurt Stockinger},
  title     = {Multiple Query Optimization Using a Gate-Based Quantum Computer},
  journal   = {IEEE Access},
  volume    = {11},
  pages     = {23066--23081},
  year      = {2023},
  doi       = {10.1109/ACCESS.2023.3254637}
}

@inproceedings{queryoptimization4,
  author    = {Nitin Nayak and Jan Rehfeld and Tobias Winker and Benjamin Warnke and Umut {\c{C}}al{\i}ky{\i}lmaz and Sven Groppe},
  title     = {Constructing Optimal Bushy Join Trees by Solving {QUBO} Problems on Quantum Hardware and Simulators},
  booktitle = {Proceedings of the International Workshop on Big Data in Emergent Distributed Environments (BiDEDE)},
  year      = {2023},
  pages     = {1--7},
  doi       = {10.1145/3579142.3594298}
}

@article{queryoptimization5,
  author    = {Manuel Sch{\"o}nberger and Stefanie Scherzinger and Wolfgang Mauerer},
  title     = {Ready to Leap (by Co-Design)? Join Order Optimisation on Quantum Hardware},
  journal   = {Proceedings of the ACM on Management of Data},
  volume    = {1},
  number    = {1},
  pages     = {1--27},
  year      = {2023},
  doi       = {10.1145/3581830}
}

@inproceedings{indexselection1,
  title     = {Index Tuning with Machine Learning on Quantum Computers for Large-Scale Database Applications},
  author    = {Gruenwald, Le and Zhang, Rui and Zhou, Jialu},
  booktitle = {Proceedings of the VLDB 2023 Workshops},
  year      = {2023},
  publisher = {VLDB Endowment},
}

@article{indexselection2,
  title   = {Index Advisors on Quantum Platforms},
  author  = {Kesarwani, Manish and Haritsa, Jayant R.},
  journal = {Proceedings of the VLDB Endowment},
  volume  = {17},
  number  = {12},
  year    = {2024},
  publisher = {VLDB Endowment}
}

@article{dataintegration1,
  title   = {Solving Hard Variants of Database Schema Matching on Quantum Computers},
  author  = {Fritsch, Kai and Scherzinger, Stefanie},
  journal = {Proceedings of the VLDB Endowment},
  volume  = {16},
  number  = {12},
  pages   = {3990--3993},
  year    = {2023},
  publisher = {VLDB Endowment}
}

@inproceedings{transactionscheduling1,
  title     = {Avoiding Blocking by Scheduling Transactions Using Quantum Annealing},
  author    = {Bittner, Tim and Groppe, Sven},
  booktitle = {Proceedings of the 24th Symposium on International Database Engineering \& Applications (IDEAS)},
  year      = {2020},
  pages     = {1--10},
  publisher = {ACM}
}

@article{transactionscheduling2,
  title   = {Hardware Accelerating the Optimization of Transaction Schedules via Quantum Annealing by Avoiding Blocking},
  author  = {Bittner, Tim and Groppe, Sven},
  journal = {Open Journal of Cloud Computing},
  volume  = {7},
  number  = {1},
  pages   = {1--21},
  year    = {2020}
}

@inproceedings{transactionscheduling3,
  title     = {Optimizing Transaction Schedules on Universal Quantum Computers via Code Generation for Grover’s Search Algorithm},
  author    = {Groppe, Sven and Groppe, Jinghua},
  booktitle = {Proceedings of the 25th International Database Engineering \& Applications Symposium (IDEAS)},
  year      = {2021},
  pages     = {149--156},
  publisher = {ACM}
}

@article{rieger2024operationalqdb,
  title   = {Operational Framework for a Quantum Database},
  author  = {Rieger, Carla and Grossi, Michele and Guerreschi, Gian Giacomo and Vallecorsa, Sofia and Werner, Martin},
  journal = {arXiv preprint arXiv:2405.14947},
  year    = {2024},
  note    = {Defines operations and implementation of quantum data storage and manipulation in a quantum database framework},
  url     = {https://arxiv.org/abs/2405.14947}
}

@inproceedings{QML1,
  author    = {Winker, Tobias and Groppe, Sven and Uotila, Valter and Yan, Zhengtong and Lu, Jiaheng and Franz, Maja and Mauerer, Wolfgang},
  title     = {Quantum Machine Learning: Foundation, New Techniques, and Opportunities for Database Research},
  booktitle = {Companion of the 2023 International Conference on Management of Data (SIGMOD/PODS)},
  year      = {2023},
  pages     = {45--52},
  publisher = {Association for Computing Machinery},
  url       = {https://doi.org/10.1145/3555041.3589404}
}

@inproceedings{QML2,
  author    = {Vogrin, Martin and Vogrin, Rok and Groppe, Sven and Groppe, Jinghua},
  title     = {Supervised Learning on Relational Databases with Quantum Graph Neural Networks},
  booktitle = {Proceedings of the Second International Workshop on Quantum Data Science and Management (QDSM’24)},
  year      = {2024},
  url       = {https://www.vldb.org/workshops/2024/proceedings/QDSM/QDSM.5.pdf}
}

@inproceedings{QML3,
  title   = {A Theoretical Framework for Learning from Quantum Data},
  author  = {Heidari, Mohsen and Padakandla, Arun and Szpankowski, Wojciech},
  booktitle = {IEEE International Symposium on Information Theory (ISIT)},
  year    = {2021},
  url     = {https://arxiv.org/abs/2107.06406},
  note    = {Theoretical foundations for learning from quantum data},
}

@inproceedings{QML4,
  title   = {Toward Physically Realizable Quantum Neural Networks},
  author  = {Heidari, Mohsen and Grama, Ananth Y. and Szpankowski, Wojciech},
  booktitle = {Association for the Advancement of Artificial Intelligence (AAAI)},
  year    = {2022},
  url     = {https://www.cs.purdue.edu/homes/spa/papers/AAAI21.pdf},
  note    = {Implementation and training scheme for quantum neural networks},
}

@article{quantum-bptree-2024,
  author       = {Hao Liu and
                  Xiaotian You and
                  Raymond Chi{-}Wing Wong},
  title        = {First Tree-like Quantum Data Structure: Quantum {B+} Tree},
  journal      = {CoRR},
  volume       = {abs/2405.20416},
  year         = {2024}
}

@inproceedings{Grover,
  author    = {Lov K. Grover},
  title     = {A fast quantum mechanical algorithm for database search},
  booktitle = {Proceedings of the 28th Annual ACM Symposium on Theory of Computing (STOC)},
  year      = {1996},
  pages     = {212--219}
}

@article{AE,
  author    = {Gilles Brassard and Peter H{\o}yer and Michele Mosca and Alain Tapp},
  title     = {Quantum Amplitude Amplification and Estimation},
  journal   = {Contemporary Mathematics},
  volume    = {305},
  pages     = {53--74},
  year      = {2002}
}

@article{DiskJoin,
  title={DiskJoin: Large-scale Vector Similarity Join with SSD},
  author={Chen, Yanqi and Yan, Xiao and Meliou, Alexandra and Lo, Eric},
  journal={ACM SIGMOD},
  volume={3},
  number={6},
  pages={1--27},
  year={2025},
  publisher={ACM New York, NY, USA}
}

@article{simjoin,
  title        = {Fast Approximate Similarity Join in Vector Databases},
  author       = {Xie, Jiadong and Yu, Jeffrey Xu and Liu, Yingfan},
  journal      = {ACM SIGMOD},
  volume       = {3},
  number       = {3},
  articleno    = {158},
  numpages     = {26},
  year         = {2025},
  month        = jun,
  doi          = {10.1145/3725403}
}

@inproceedings{simjoin2,
  title={Implementing Distributed Approximate Similarity Joins using Locality Sensitive Hashing},
  author={Aum{\"u}ller, Martin and Ceccarello, Matteo},
  booktitle={EDBT: International Conference on Extending Database Technology},
  year={2022}
}

@article{simjoin3,
  title={Efficient continuous kNN join over dynamic high-dimensional data},
  author={Ukey, Nimish and Zhang, Guangjian and Yang, Zhengyi and Li, Binghao and Li, Wei and Zhang, Wenjie},
  journal={World Wide Web},
  volume={26},
  number={6},
  pages={3759--3794},
  year={2023},
  publisher={Springer}
}

@inproceedings{vldb02,
  title={Generic database cost models for hierarchical memory systems},
  author={Manegold, Stefan and Boncz, Peter and Kersten, Martin L},
  booktitle={VLDB'02: Proceedings of the 28th International Conference on Very Large Databases},
  pages={191--202},
  year={2002},
  organization={Elsevier}
}

@inproceedings{durr1996quantum,
  title     = {A Quantum Algorithm for Finding the Minimum},
  author    = {D{\"u}rr, Christoph and H{\o}yer, Peter},
  booktitle = {Proceedings of the 5th International Conference on Quantum Computing and Quantum Communications (QCQC)},
  pages     = {362--371},
  year      = {1996},
  publisher = {Springer, Berlin, Heidelberg}
}

@article{quanto,
  author       = {Jessica Pointing and
                  Oded Padon and
                  Zhihao Jia and
                  Henry Ma and
                  Auguste Hirth and
                  Jens Palsberg and
                  Alex Aiken},
  title        = {Quanto: Optimizing Quantum Circuits with Automatic Generation of Circuit
                  Identities},
  journal      = {CoRR},
  volume       = {abs/2111.11387},
  year         = {2021}
}

@article{DBLP:journals/qip/Shyamsundar23,
  author       = {Prasanth Shyamsundar},
  title        = {Non-Boolean quantum amplitude amplification and quantum mean estimation},
  journal      = {Quantum Inf. Process.},
  volume       = {22},
  number       = {12},
  pages        = {423},
  year         = {2023}
}

@inproceedings{qir,
  author       = {Yannick Stade and
                  Lukas Burgholzer and
                  Robert Wille},
  title        = {Towards Supporting {QIR:} Steps for Adopting the Quantum Intermediate
                  Representation},
  booktitle    = {{SC} Workshops},
  pages        = {1907--1915},
  publisher    = {{ACM}},
  year         = {2025}
}

@inproceedings{SWAPTEST,
  author       = {Wang Fang and
                  Qisheng Wang},
  title        = {Optimal Quantum Algorithm for Estimating Fidelity to a Pure State},
  booktitle    = {{ESA}},
  series       = {LIPIcs},
  volume       = {351},
  pages        = {4:1--4:12},
  publisher    = {Schloss Dagstuhl - Leibniz-Zentrum f{\"{u}}r Informatik},
  year         = {2025}
}

@inproceedings{datasearch2,
  author       = {Lov K. Grover},
  title        = {A Fast Quantum Mechanical Algorithm for Database Search},
  booktitle    = {{STOC}},
  pages        = {212--219},
  publisher    = {{ACM}},
  year         = {1996}
}

@article{DBLP:journals/pvldb/ZengHSPMZ23,
  author       = {Xinyu Zeng and
                  Yulong Hui and
                  Jiahong Shen and
                  Andrew Pavlo and
                  Wes McKinney and
                  Huanchen Zhang},
  title        = {An Empirical Evaluation of Columnar Storage Formats},
  journal      = {Proc. {VLDB} Endow.},
  volume       = {17},
  number       = {2},
  pages        = {148--161},
  year         = {2023}
}

@book{singlequbit1,
  author       = {Michael A. Nielsen and
                  Isaac L. Chuang},
  title        = {Quantum Computation and Quantum Information (10th Anniversary edition)},
  publisher    = {Cambridge University Press},
  year         = {2016}
}

@article{singlequbit2,
  author       = {Mikko M{\"{o}}tt{\"{o}}nen and
                  Juha J. Vartiainen and
                  Ville Bergholm and
                  Martti M. Salomaa},
  title        = {Transformation of quantum states using uniformly controlled rotations},
  journal      = {Quantum Inf. Comput.},
  volume       = {5},
  number       = {6},
  pages        = {467--473},
  year         = {2005}
}

@inproceedings{multiqubitgate1,
  author       = {D. Michael Miller and
                  Robert Wille and
                  Zahra Sasanian},
  title        = {Elementary Quantum Gate Realizations for Multiple-Control Toffoli
                  Gates},
  booktitle    = {{ISMVL}},
  pages        = {288--293},
  publisher    = {{IEEE} Computer Society},
  year         = {2011}
}

@article{largecandidate,
  author       = {Pablo Arrighi and
                  Gilles Dowek},
  title        = {The Physical Church-Turing Thesis and the Principles of Quantum Theory},
  journal      = {Int. J. Found. Comput. Sci.},
  volume       = {23},
  number       = {5},
  pages        = {1131--1146},
  year         = {2012}
}

@article{quantumANN,
  author       = {Nathan Wiebe and
                  Ashish Kapoor and
                  Krysta M. Svore},
  title        = {Quantum algorithms for nearest-neighbor methods for supervised and
                  unsupervised learning},
  journal      = {Quantum Inf. Comput.},
  volume       = {15},
  number       = {3{\&}4},
  pages        = {316--356},
  year         = {2015}
}

@online{QuteLink,
  author    = {},
  title     = {(Quantum Database Prototype)},
  year      = {Qute},
  url       = {https://github.com/weAIDB/Qute}
}



\end{document}